# Radiation Heat Transfer in Particle-Laden Gaseous Flame: Flame Acceleration and Triggering Detonation

M. A. Liberman[1][*)]  M. F Ivanov[2], A. D. Kiverin,[2]

[1] Nordita, KTH Royal Institute of Technology and Stockholm University
Stockholm, Sweden
[2] Joint Institute for High Temperatures of RAS, Moscow, Russia

27 March 2015

**Abstract**

In this study we examine influence of the radiation heat transfer on the combustion regimes in the mixture, formed by suspension of fine inert particles in hydrogen gas. The gaseous phase is assumed to be transparent for the thermal radiation, while the radiant heat absorbed by the particles is then lost by conduction to the surrounding gas. The particles and gas ahead of the flame is assumed to be heated by radiation from the original flame. It is shown that the maximum temperature increase due to the radiation preheating becomes larger for a flame with lower velocity. For a flame with small enough velocity temperature of the radiation preheating may exceed the crossover temperature, so that the radiation heat transfer may become a dominant mechanism of the flame propagation. In the case of non-uniform distribution of particles, the temperature gradient formed due to the radiation preheating can initiate either deflagration or detonation ahead of the original flame via the Zel'dovich's gradient mechanism. The initiated combustion regime ignited in the preheat zone ahead of the flame depends on the radiation absorption length and on the steepness of the formed temperature gradient. Scenario of the detonation triggering via the temperature gradient mechanism formed due to the radiation preheating is plausible explanation of the transition to detonation in Supernovae Type Ia explosion.

[*)] Corresponding author at Nordic Institute for Theoretical Physics (NORDITA): Tel: +46855378444
E-mail address: mliber@nordita.org  (M. A. Liberman)

# 1. Introduction

Notoriously, while studying combustion in gaseous mixture, the radiation of hot combustion products is usually not important, as the radiation absorption length in a gaseous mixture is very large, so that the gaseous mixture is almost fully transparent for the radiation and therefore the radiation heat transfer does not influence the flame dynamics. For example, the photon mean free path in the atmosphere at pressure $P = 1 atm$ is about tens meters because of very small $(10^{-24} \div 10^{-25}) cm^2$ values of the Thompson scattering and "bremsstrahlung" cross section processes. Therefore the contribution of radiation to the heat transfer is negligibly small. If the flame propagates in a tube out from the closed to the open end, the radiation heat losses of the hot combustion products cause a relatively modest cooling of the burned products resulting in a modest decrease of pressure behind the flame front, which is negligible compared to the thermal conduction heat losses to the tube walls. In the traditional theoretical combustion the heat is transferred by the molecular gaseous thermal conduction and/or convection while the radiation heat transfer is negligible because the energy transferred by the radiant heat flux contributes far too small to the mechanism of combustion wave propagation and does not influence the flame velocity.

The situation changes drastically if the gaseous mixture is seeded with fine inert particles, which absorbed and heated by thermal radiation and then transfer the heat by conduction to the surrounding gas. In this case the gas temperature ahead of the flame lags that of the particles and the radiation preheating causes either acceleration of the flame or non-uniform temperature distribution with a proper temperature gradient, formed ahead of the flame, trigger either new deflagration or detonation via the Zel'dovich gradient mechanism. In the present paper we investigate the influence of the radiation preheating for the particle-laden hydrogen-oxygen and hydrogen/air flames. Scenario of the radiation preheating resulting in the triggering detonation can be plausible mechanism explaining deflagration-to-detonation transition in the thermonuclear Type Ia supernovae (SN Ia), which still remains the least understood aspect of the SN Ia explosion phenomenon. In the case of non-uniform distribution of particles, which is typical e.g. for dust deposits layers ("methane-air detonation" in coal dust), the time of the radiative heating is longer. If time of the flame arrival to the boundary of denser particles layer, where the radiation is noticeably absorbed, is long enough, the maximum temperature within the temperature gradient established due to the radiative preheating may exceed the crossover value. In this case either new deflagration or detonation can be ignited via the Zel'dovich's gradient mechanism [1, 2]. What kind of combustion regime is ignited in the "distant" particle seeded cloud depends on the radiation absorption length and on the steepness of the formed temperature gradient.

It is known, that uncontrolled development of detonation poses significant threats to chemical storage and processing facilities, mining operations, etc. [3-5], while controlled detonation initiation can be a potential application for propulsion and power devices [6, 7]. Detonations may or may not develop depending on the ability of a particular mixture composition to sustain detonations, and on the ability of flames to accelerate and produce shocks that are strong enough to ignite detonation. Study of the premixed flames and detonations arising and propagating in the particle-laden gaseous combustible mixture is important for the understanding of unconfined vapor cloud explosions and accidental explosions in many industrial processes associated with the risk of dust explosions, and for better performance of rocket engines using fluid or solid fuels, see e.g. [8, 9]..

Majority of the previous studies used a one-step chemical reaction model and were mainly focused on the deflagration-to-detonation transition (DDT) in a gaseous combustible mixtures in attempt to understand nature of the detonation formation. Although significant progress has been made in the understanding of the flame dynamics, the nature of the transition to detonation still remained highly uncertain because a one-step chemical model allows ignition at any temperatures, so that the results of such studies were often remained questionable. In the 1980s-1990s, several groups used a one-step Arrhenius chemical model and asymptotic methods for high activation energy to examine effects of the radiation on the flames propagating in a gas mixture seeded by solid particles [10, 11]. The flame propagating in the presence of the uniformly dispersed inert solid particles has been considered with and without account of radiative heat transfer [12-18]. Coal combustion research has been focused mainly on two aspects of practical interest: the production of volatiles due to thermal decomposition of coal dust and char combustion [19-21]. The combustible volatiles can react and release energy, which in turn may contribute to the heat-up of the particles, enhance the combustion energy release due to energy feedback mechanism resulting in an explosion. For the coal-dust suspension air filling the coal-fired burners and for rocket engines using the solid or fluid fuels as well as for coal-fire mining safety both the ignition and combustion evolution are of paramount importance. Effect of radiation transfer on a spray combustion can be of interest for practical cases such as diesel engines, gas turbine combustors etc.

A combustible mixture can be ignited by electrical sparks, or by thermal heating. The ignition capability of an electrical spark varies with fuel concentration, humidity, oxygen content of the atmosphere, temperature, and turbulence, requiring about 0.01-0.03mJ depending on the mixture reactivity. In contrast, radiation-induced ignition typically requires much larger amounts of energy to be released in the mixture. Direct thermal ignition of gaseous combustible mixture by absorption of radiation causing a rapid increase in temperature at least up to 1000K is possible by focusing a high power laser radiation and has been demonstrated both theoretically and experimentally [22, 23]. However, ignition at low power levels is unlikely because of a very large length of absorption of the combustible gases at normal conditions.

In the present study effects of thermal radiative preheating is considered for the flames propagating in a two phases composite comprising of gaseous combustible mixture and inert particles. Recent experiments have shown that the dust cloud flame propagation is strongly influenced by the thermal radiation [24-27]. The effects of the radiation preheating is investigated for the hydrogen-oxygen and hydrogen-air flame. The gaseous phase is assumed to be transparent for the radiation, while solid particles absorb and reemit the radiation. Different scenarios are considered depending on spatial distribution of the suspended particles and the laminar flame velocity in a pure gaseous mixture. It the case of uniform spatial distribution of the particles the thermal radiation emitted from the hot combustion products is absorbed by the particles ahead of the flame resulting in the radiation preheating, which in turn causes the increase of the flame velocity. It is shown also that the radiative heat flux from the primary particle-laden flame may generate secondary explosion ahead of the flame in the distant particle cloud. This phenomenon is demonstrated for the non-uniform spatial distribution of particles, when the radiation absorbed far ahead of the flame creates a nonuniform temperature distribution in the unburned mixture. If maximum temperature ahead of the flame rises up to the crossover value before the flame arrival to this location, then either new deflagration or detonation can be ignited via the Zeldovich gradient mechanism.

The paper is organized as follows. In Section 2 we present the mathematical model used to study the problem in question. A simple model explaining the principal features of the radiation preheating, and numerical study of the influence of the radiative preheating on combustion wave velocity for a uniform spatial distribution of the particles is presented in Section 3. Triggering of deflagration or detonation ahead of the flame depending on the steepness of temperature gradient created by the radiative preheating in the case of non-uniform particle distribution is considered in Section 4. We conclude in Sec. 5. Appendix A presents validation and thorough convergence and resolution tests of the numerical scheme used in the present studies.

## 2. Formulation of the problem; Governing equations

We consider a planar flame propagating from the closed to the open end of a duct. The governing equations for the gaseous phase are the one-dimensional, time-dependent, multispecies reactive Navier-Stokes equations including the effects of compressibility, molecular diffusion, thermal conduction, viscosity and detailed chemical kinetics for the reactive species $H_2$, $O_2$, $H$, $O$, $OH$, $H_2O$, $H_2O_2$, and $HO_2$ with subsequent chain branching, production of radicals, energy release and heat transfer between the particles and the gas phase. The system of equations for the gaseous phase is

$$\frac{\partial \rho}{\partial t} + \frac{\partial (\rho u)}{\partial x} = 0, \tag{1}$$

$$\frac{\partial Y_i}{\partial t} + u\frac{\partial Y_i}{\partial x} = \frac{1}{\rho}\frac{\partial}{\partial x}\left(\rho D_i \frac{\partial Y_i}{\partial x}\right) + \left(\frac{\partial Y_i}{\partial t}\right)_{ch}, \tag{2}$$

$$\rho\left(\frac{\partial u}{\partial t} + u\frac{\partial u}{\partial x}\right) = -\frac{\partial P}{\partial x} + \frac{\partial \sigma_{xx}}{\partial x} - \rho_p \frac{(u-u_p)}{\tau_{St}}, \tag{3}$$

$$\rho\left(\frac{\partial E}{\partial t} + u\frac{\partial E}{\partial x}\right) = -\frac{\partial (Pu)}{\partial x} + \frac{\partial}{\partial x}(\sigma_{xx} u) + \frac{\partial}{\partial x}\left(\kappa(T)\frac{\partial T}{\partial x}\right) +$$

$$+\sum_k h_k\left(\frac{\partial}{\partial x}\left(\rho D_k(T)\frac{\partial Y_k}{\partial x}\right)\right) + \rho\sum_k h_k\left(\frac{\partial Y_k}{\partial t}\right)_{ch} - \rho_p u_p \frac{(u-u_p)}{\tau_{St}} - \rho_p c_{P,p} Q, \tag{4}$$

$$P = R_B T n = \left(\sum_i \frac{R_B}{m_i} Y_i\right)\rho T = \rho T \sum_i R_i Y_i, \tag{5}$$

$$\varepsilon = c_v T + \sum_k \frac{h_k \rho_k}{\rho} = c_v T + \sum_k h_k Y_k, \tag{6}$$

$$\sigma_{xx} = \frac{4}{3}\mu\left(\frac{\partial u}{\partial x}\right), \tag{7}$$

The standard notations are use: $P$, $\rho$, $u$, are pressure, mass density, and flow velocity of the gaseous mixture, $Y_i = \rho_i/\rho$ - the mass fractions of the species, $E = \varepsilon + u^2/2$ - the total energy density, $\varepsilon$ - the internal energy density, $R_B$ - is the universal gas constant, $m_i$ - the molar mass of i-species, $R_i = R_B/m_i$, n - the gaseous molar density, $\sigma_{ij}$ - the viscous stress tensor, $c_v = \sum_i c_{vi} Y_i$ - is the constant volume specific heat, $c_{vi}$ - the constant volume specific heat of i-species, $h_i$ - the enthalpy of formation of i-species, $\kappa(T)$ and $\mu(T)$ are the coefficients of thermal conductivity and viscosity, $D_i(T)$ - is the diffusion coefficients of i-species, $(\partial Y_i/\partial t)_{ch}$ - is the variation of i-species concentration (mass fraction) in chemical reactions, $\rho_p = m_p N_p$ is mass density of the suspended particles, $N_p$ – the particles number density, $u_p$, $r_p$, $m_p$ – velocity, radius and mass of the spherical particle, $\tau_{St} = m_p/6\pi\mu r_p$ - the Stokes time, $Q$ – interphase thermal exchange source, $c_{P,p}$ – the constant pressure specific heat of the particle material.

The changes in concentrations of the mixture components due in the chemical reactions are defined by the solution of system of chemical kinetics

$$\frac{dY_i}{dt} = F_i(Y_1, Y_2, ...Y_N, T), \quad i = 1, 2, ...N. \tag{8}$$

The right hand parts of Eq. (8) contain the rates of chemical reactions for the reactive species. We use the standard reduced chemical kinetic scheme for hydrogen/oxygen combustion with the elementary reactions of the Arrhenius type and with pre-exponential constants and activation energies presented in [28]. This reaction scheme for a stoichiometric $H_2/O_2$ mixture has been tested in many applications and to a large extent adequate to complete chemical kinetics well describing the main features of the $H_2/O_2$ combustion. The computed thermodynamic, chemical, and material parameters using this chemical scheme are in a good agreement with the flame and detonation parameters measured experimentally. For example, for $P_0 = 1.0\,\text{bar}$ we obtain for the laminar flame velocity, the flame thickness and adiabatic flame temperature $U_f \approx 12\,\text{m/s}$, $L_f = 0.24\,\text{mm}$, $T_b = 3012\,\text{K}$, correspondingly, for the expansion ratio (the ratio of the density of unburned gas and the combustion products) $\Theta = \rho_u / \rho_b = 8.36$, and for temperature and velocity of CJ-detonation $T_{CJ} = 3590\,\text{K}$, $U_{CJ} = 2815\,\text{m/s}$.

The equations of state of the fresh mixture and combustion products are taken with the temperature dependence of the specific heats, heat capacities and enthalpies of each species borrowed from the JANAF tables and interpolated by the fifth-order polynomials [29]. The transport coefficients were calculated using the gas kinetics theory [30]. The gaseous mixture viscosity coefficients are

$$\mu = \frac{1}{2}\left[\sum_i \alpha_i \mu_i + \left(\sum_i \frac{\alpha_i}{\mu_i}\right)^{-1}\right], \qquad (9)$$

where $\alpha_i = \dfrac{n_i}{n}$ is the molar fraction, $\mu_i = \dfrac{5}{16}\dfrac{\sqrt{\pi \hat{m}_i kT}}{\pi \Sigma_i^2 \tilde{\Omega}_i^{(2,2)}}$ is the viscosity coefficient of $i$-species, $\tilde{\Omega}^{(2,2)}$ - is the collision integral which is calculated using the Lennard-Jones potential [30], $\hat{m}_i$ is the molecule mass of the i-th species of the mixture, $\Sigma_i$ is the effective molecule size. The thermal conductivity coefficient of the gas mixture is

$$\kappa = \frac{1}{2}\left[\sum_i \alpha_i \kappa_i + \left(\sum_i \frac{\alpha_i}{\kappa_i}\right)^{-1}\right]. \qquad (10)$$

Coefficient of the heat conduction of i-th species $\kappa_i = \mu_i c_{pi} / \Pr$ can be expressed via the kinematic viscosity $\mu_i$ and the Prandtl number, which is taken $\Pr \approx 0.71 \div 0.75$.

The binary coefficients of diffusion are

$$D_{ij} = \frac{3}{8} \frac{\sqrt{2\pi kT \hat{m}_i \hat{m}_j / (\hat{m}_i + \hat{m}_j)}}{\pi \Sigma_{ij}^2 \tilde{\Omega}^{(1,1)}(T_{ij}^*)} \cdot \frac{1}{\rho}, \qquad (11)$$

where $\Sigma_{ij} = 0,5(\Sigma_i + \Sigma_j)$, $T_{ij}^* = kT/\varepsilon_{ij}^*$, $\varepsilon_{ij}^* = \sqrt{\varepsilon_i^* \varepsilon_j^*}$; $\varepsilon^*$ are the constants in the expression of the Lennard-Jones potential, and $\tilde{\Omega}_{ij}^{(1,1)}$ is the collision integral similar to $\tilde{\Omega}^{(2,2)}$ [30]. The diffusion coefficient of i-th species is

$$D_i = (1 - C_i) / \sum_{i \neq j} \alpha_i / D_{ij}. \qquad (12)$$

A detailed description of the transport coefficients used for the gaseous phase and calculation of diffusion coefficients for intermediates has been published previously, see e.g. [31-33].

The dynamics of solid particles is considered in continuous hydrodynamic approximation. The interaction between particles is assumed negligibly small for a small volumetric concentration of particles, so that only the Stokes force between the particle and gaseous phase is taken into account. Then, the equations for the phase of suspended particles are:

$$\frac{\partial N_p}{\partial t} + \frac{\partial (N_p u_p)}{\partial x} = 0 \qquad (13)$$

$$\left( \frac{\partial u_p}{\partial t} + u_p \frac{\partial u_p}{\partial x} \right) = \frac{(u - u_p)}{\tau_{St}}, \qquad (14)$$

$$\left( \frac{\partial T_p}{\partial t} + u_p \frac{\partial T_p}{\partial x} \right) = Q - \frac{2\pi r_p^2 N_p}{c_{P,p} \rho_{p0}} \left( 4\sigma T_p^4 - q_{rad} \right). \qquad (15)$$

Where $T_p$ – temperature of the particles, $2\pi r_p^2 N_p \left( 4\sigma T_p^4 - q_{rad} \right)$ - is the thermal radiation heat flux absorbed and reemitted by the particles. The heat transferred from the particle surface to surrounding gaseous mixture is

$$Q = (T_p - T) / \tau_{pg}, \qquad (16)$$

where $\tau_{pg} = 2r_p^2 c_{P,p} \rho_{p0} / 3\kappa Nu$ is the characteristic times of the energy transfer from the particle surface to surrounding gaseous mixture, $c_{P,p}$ and $\rho_{p0}$ are specific heat and the mass density of the particle material, Nu is the Nusselt number (see e.g. [34]).

For a one-dimensional planar problem the equation for the thermal radiation heat transfer in the diffusion approximation is [35, 36]:

$$\frac{d}{dx}\left(\frac{1}{\chi}\frac{dq_{rad}}{dx}\right) = -3\chi\left(4\sigma T_p^4 - q_{rad}\right), \qquad (17)$$

where the radiation absorption coefficient is $\chi = 1/L = \pi r_p^2 N_p$, and $L = 1/\pi r_p^2 N_p$ is the radiation absorption length. The particle-laden mixture is considered optically-thick so that the radiation energy flux emitted from the flame, which is mainly determined by the particle concentration is assumed to be equal to the blackbody radiative heat source, $q_{rad}(x = X_f) \approx \sigma T_b^4$, where $\sigma = 5.6703 \cdot 10^{-8} \, W/m^2 K^4$ is the Stefan-Boltzmann constant, and $T_b$ is temperature of the combustion products. The calculations were carried out for stoichiometric hydrogen-oxygen and hydrogen-air mixtures at initial pressure $P_0 = 1 atm$ with a small solid inert spherical particles suspended in the gaseous mixture. For simplicity the particles are assumed to be identical with the mass density of the particles, $\rho_p = m_p N_p$ much smaller than the gas density, $\zeta = m_p N_p / \rho \ll 1$, so that there is only one way of a momentum coupling of the particles and the gaseous phase.

The numerical method used in the present studies is based on splitting of the Eulerian and Lagrangian stages, known as Coarse Particle Method (CPM). For the first time it has been developed by Gentry, Martin and Daly [37] and afterwards was modified and widely implemented by Belotserkovsky and Davydov [38]. It was further modified [39] so that a high numerical stability of the method is achieved if the hydrodynamic variables are transferred across the grid boundary with the velocity, which is an average value of the velocities in neighboring grids. The improvement of overall modified solver implemented in [39] provides the second order in space that differs from the original first-order method [38]. The modified CPM solver was thoroughly tested and successfully used for modeling knock appearance in SI-engines [39, 40] and to study the flame acceleration in tubes with non-slip walls, the transition form slow combustion to detonation [31-33] and other problems of transient combustion, e.g. ignition of different combustion modes [41-43]. The system of chemical kinetics equations represents a stiff system of differential equations, and it was solved using standard Gear's method [44]. The developed algorithm was implemented using the FORTRAN-90. Convergence and resolution tests presented in Appendixes A were carried out to verify that the observed phenomena are correctly caught remaining unchanged with increasing resolution.

**3. Radiation heat transfer and flame propagation for uniformly dispersed particles**

We consider the hydrogen-oxygen flame propagating in the mixture with uniformly dispersed identical solid spherical particles of radius $r_p = 0.75 \mu m$, density of the particle material, $\rho_{p,0} = 1 g/cm^3$, and concentrations $N_p = 5.7 \cdot 10^7; 2.85 \cdot 10^7; 1.4 \cdot 10^7 cm^{-3}$, that corresponds to the radiation absorption lengths: $L = 1, 2,$ and $4 cm$. As it was mentioned above, the particles ahead of the flame absorb the thermal radiation, their temperature increases and they transfer the heat to the surrounding gas mixture. The characteristic time scales of the problem for the chosen parameters are $\tau_{St} = 2r_p^2 \rho_{p0} / 9\nu_g \rho_g \approx 3\mu s$, $\tau_{pg} \approx 1\mu s$; the characteristic gas-dynamic time scales are $L_f / U_f \approx 20 \mu s$ and $L / U_f \approx 1 ms$. Since the characteristic time of energy transfer between the particles and the gaseous phase $\tau_{pg}$ is much smaller than the characteristic gas-dynamic time scales, the temperatures of the particles and the gaseous phase are approximately equal, $T_p \approx T$. The mass loading parameter values are $\varsigma = \rho_p / \rho_g = 0.2$; 0.1 and 0.05 for $L = 1, 2$ and 4cm, correspondingly. Because $\varsigma = \rho_p / \rho_g \ll 1$, the momentum coupling of the particles and the gaseous phase is small, and also is small the influence of the particles on the flame dynamics. Thus, only the heating of particles by the absorbed radiation and corresponding heating of the gas mixture ahead of the flame will influence the flame dynamics. The presence of neutral solid particles is similar to the dilution of combustible mixture with inert gas decreasing the adiabatic temperature behind the flame. This effect is also small for $\varsigma(c_{p,p} / c_{V,g}) \ll 1$. This choice of the parameters allows us to distinguish the effect only of the radiation preheating on the flame dynamics.

Taken into account that the stationary flow is established during the time of the order $L / U_f$, it is straightforward to obtain an estimate for the maximum temperature increase ahead of the flame due to the radiation preheating in the laden-particle mixture. In the coordinate system co-moving with the flame front the unburned mixture with suspended particles flows toward the flame with the normal laminar flame velocity $U_f$. The thermal radiation is appreciably absorbed by the particles, which are located at $x \leq L$ ahead of the flame front, and their temperature can be estimated from the energy balance

$$\rho_p c_{p,p} \frac{dT_p}{dt} = \sigma T_b^4 \exp\left(-\frac{x - U_f t}{L}\right) \pi r_p^2 N_p - \frac{\rho_p c_{p,p}}{\tau_{pg}} (T_p - T). \qquad (18)$$

The gaseous phase temperature ahead of the flame increases due to the heat transferred from the particles to the surrounding unburned mixture

$$\frac{dT}{dt} = -\frac{(T - T_p)}{\tau_{gp}} \qquad (19)$$

Taken into account that $T \approx T_p$, and $L = 1 / \pi r_p^2 N_p$, we obtain from (18) and (19)

$$\rho_p c_{p,p} (1 + \xi) \frac{dT}{dt} = \sigma T_b^4 \frac{1}{L} \exp\left(-\frac{x - U_f t}{L}\right), \qquad (20)$$

where $\xi = \tau_{gp} / \tau_{pg} = c_{V,g} / \varsigma c_{P,p}$.

The characteristic time of the radiation preheating of the Lagrangian particle is approximately the time of its arrival to the flame front, $t \approx L / U_f$. Taking this into account, the maximum temperature increase of the unburned mixture close ahead of the flame can be estimated as:

$$\Delta T = \sigma T_b^4 \frac{1}{U_f} \frac{(1-e^{-1})}{\rho_p c_p (1+\xi)} \approx 0.63 \frac{\sigma T_b^4}{(\rho_p c_p + \rho c_{V,g}) U_f}. \tag{21}$$

It can be readily observed that the maximum temperature increase due to the radiation preheating does not depend on the radiation absorption length (see Fig.1). This is due to the fact that although the local radiant heat flux absorbed by the particle is larger for smaller absorption length, but for a larger absorption length particles absorb the radiant heat flux over a longer time until their arrival to the flame front. This difference in time for smaller and larger absorption lengths compensates the lesser local heating for a larger absorption lengths.

Taking into account that the adiabatic flame temperature is less than it is in a pure mixture due to dilution by the inert solid particles, and that the photons emitted from the flame front are produced within the radiative layer near the flame front of the finite thickness, the effective temperature of the radiation emitted from the flame surface can be estimated as $T_{b,eff} \approx 2700K$. Then, the maximum temperature increase caused by the radiation preheating can be estimated as $\Delta T \approx 160K$, which is in a good agreement with the numerical simulation shown in Fig.1 and Fig.2. Fig. 1 shows temporal evolution of the maximum gaseous temperature increase at the distance 2mm ahead of the flame front calculated for the radiation absorption lengths $L = 1, 2$ and 4cm. One can see that the stationary values of temperature in the gaseous mixture ahead of the flame are established during $t \approx L / U_{f0}$, where $U_{f0}$ is the normal laminar flame velocity in pure gas mixture. Fig. 2 shows the profiles of the gaseous phase temperature, which are established after the stationary flow was settled, calculated for the radiation absorption lengths, L=1, 2, and 4cm.

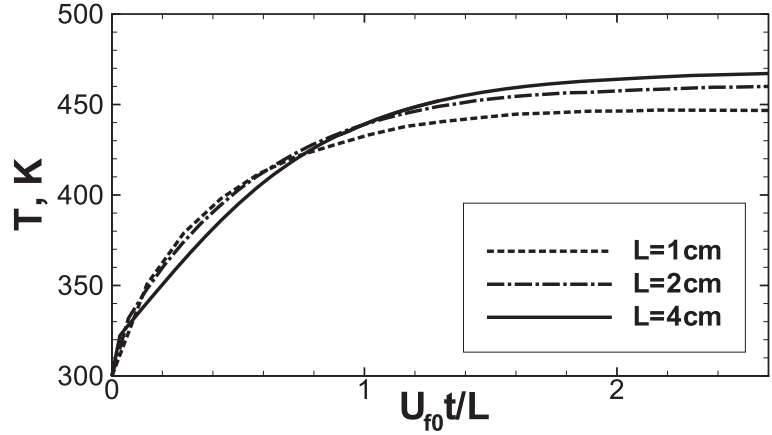

**Figure 1.** Time evolution of the gaseous temperature during radiative preheating at the distance 2mm ahead the flame front for different radiation absorption lengths, L=1, 2, and 4cm. On the x-axis time is in units $L/U_{f0}$.

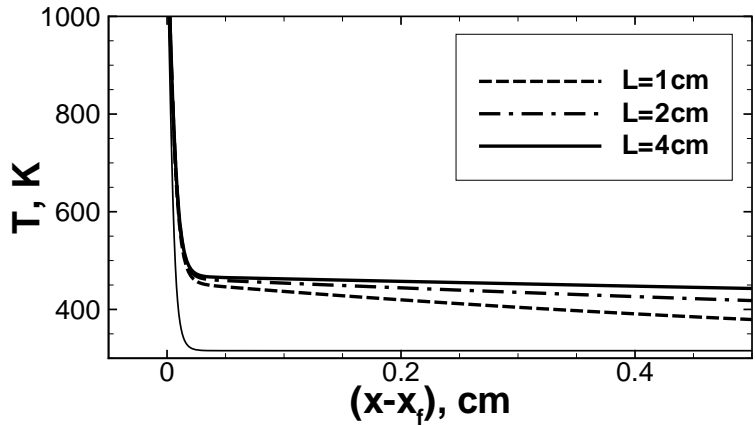

**Figure 2.** Temperature distribution ahead of the flame front calculated for uniformly distributed suspended particles and for different radiation absorption lengths, L=1, 2, and 4 cm. Thin line corresponds to the temperature profile for the laminar flame in a pure stoichiometric hydrogen/oxygen mixture.

Absorption of the thermal radiation emitted from the hot combustion products by the particles ahead of the flame results in the radiation preheating, which in turn results in the increase of the flame velocity. Fig. 3 shows the increase of the flame velocity relative to the unburned mixture due to the radiation preheating of the gaseous mixture ahead of the flame calculated for the radiation absorption lengths L=1, 2 and 4cm. Recall, that when flame propagates from the closed end of a duct, the unburned gas ahead of the flame moves to the open end with the velocity $u = (\Theta-1)U_f$, where $\Theta = \rho_u/\rho_b$ is the density ratio of the unburned $\rho_u$ and burned $\rho_b$ fuel, respectively [45, 46]. The velocity of the flame with respect to the tube walls is $U_{fL} = \Theta U_f$ and with respect to the unburned gas the flame velocity it is $U_f$. Due to the radiative preheating of the reacting gaseous mixture ahead of the flame, the flame velocity increases. Asymptotically, the flame velocity goes to the normal laminar flame velocity for $L \to \infty$, which corresponds to the pure gaseous mixture. The applicability of the model limits values of L from below, $L > 0.5 cm$, which follows from inequality $\varsigma \ll 1$ written in the form $L \gg r_p (\rho_{p0}/\rho_g)$.

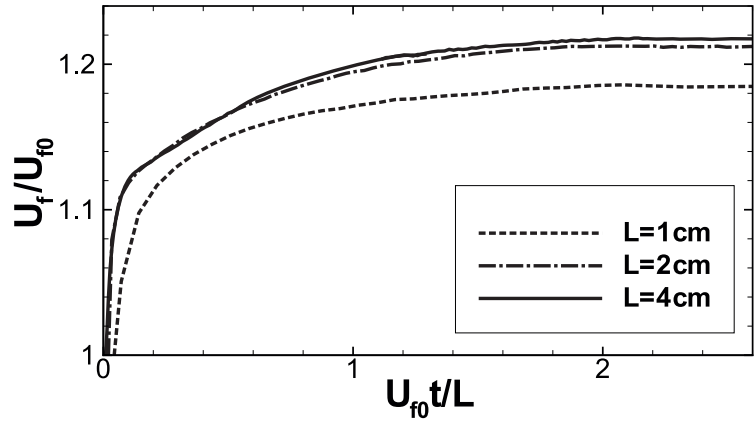

**Figure 3.** Temporal evolution of the flame velocity for the flame propagating through the gas-particles cloud with uniformly suspended micro-particles for different thermal radiation absorption lengths L. The flame velocity is normalized on the normal laminar flame velocity $U_{f0}$ in pure gas mixture. Time is in units $L/U_{f0}$.

The radiation preheating time and the temperature increase ahead of the flame attained due to the radiation preheating is larger for lower flame velocities and/or for a lower initial pressures. For a small enough velocity of the flame (as well for smaller gaseous phase density) the maximum temperature due to the radiation preheating may exceed the crossover temperature, when the endothermic reaction stage passes to the fast exothermic stage. In this case the radiative heat transfer may dominate the gaseous thermal conduction mechanism of the flame propagation. In the approximation of radiation thermal conductivity, the coefficient of radiation thermal conductivity is defined as

$$\kappa_{Rad} = 16\sigma T^3 L / 3 . \tag{22}$$

The velocity and structure of a laminar flame in the classical theory of combustion by Zel'dovich and Frank-Kamenetskii [45] is defined in the approximation of the linear thermal conduction equation. For the one-dimensional problem the temperature distribution (for constant thermal diffusivity) is

$$T(x,t) \propto \frac{1}{2\sqrt{\pi \chi_g t}} \exp(-x^2 / 4\chi_g t) , \tag{23}$$

where $\chi_g = \kappa_g / \rho_g c_{g,p} \approx \text{const}$ is the coefficient of gaseous thermal diffusivity. It follows from Eq. (23) that heat propagates at the distance $x \propto \sqrt{4\chi_g t}$. Taking this into account, it is straightforward to obtain an order-of-magnitude estimates for the speed and width of the laminar flame, which propagates as a result of thermal conduction heat transfer:

$$L_f \propto \sqrt{\chi_g \tau_R} , \quad U_f \propto \sqrt{\chi_g / \tau_R} . \tag{24}$$

This estimate is based on the consideration that the characteristic heat diffusion time scale is much longer than the characteristic time of the heat release in the reaction $\tau_R$, otherwise any small disturbances of the combustion wave will diffuse away resulting in the flame extinguishing.

Another well known conclusion of the classical combustion theory is that in the approximation of linear thermal conduction and if the coefficient of thermal diffusivity is equal to the coefficient of diffusion of reagents, the total enthalpy is constant inside of the flame, which implies similarity of the temperature and the density (concentration of reagent) distributions in the flame front structure. Contrary to the classical theory if the radiative heat transfer becomes dominating process, the heat propagation is defined by the nonlinear heat conduction equation. The total enthalpy of the unburned mixture and combustion products remains equal but the total enthalpy is not constant inside of the flame and there is no similarity of the temperature and the density distributions. In the case of dominating radiation heat transfer the temperature distribution is considerably different compared to that given by Eq. (23). The temperature distribution in the planar thermal wave for the nonlinear radiation thermal conductivity (22) is

$$T(x) \propto T_0 \left(1 - x^2 / x_0^2\right)^{1/3}. \qquad (25)$$

In this case the reaction can be ignited within the preheat zone, which size by the order of magnitude is the radiation absorption length L. Correspondingly, the flame thickness is by the order-of-magnitude equal to the radiation absorption length, i.e. $L_{fR} \propto L$, and the flame velocity can be estimated as $U_{f,rad} \propto (L/\tau_R) \approx (L/L_{f0})U_{f0} \gg U_{f0}$, where $L_{f0}$ and $U_{f0}$ are the width and velocity of the laminar flame in pure gaseous mixture, respectively. It is clear that the flame velocity for the dominating radiative heat transfer significantly exceeds the laminar flame velocity in a pure gas mixture. As soon as combustion has been initiated by the primary particle-laden flame, the combustion generates secondary explosions ahead of the flame in the particle-laden mixture, and such combustion will look like a sequence of thermal explosions, and can be accompanied by a strong increase in pressure. Presumably such scenario is likely to occur in coal dust explosion where combustible volatiles can react and release energy, which in turn may contribute to the heat-up of the particle and combustible volatiles can react and release energy, which enhance the energy release.

As an example of the influence of the radiation preheating for the flame with lower laminar velocity in a pure gas mixture, we consider the hydrogen/air flame for the same parameters of the uniformly dispersed particles as in Fig. 1. In this case the radiation preheating time is longer, because the flame velocity is approximately 6 times smaller than for the hydrogen oxygen flame. Because of the lower adiabatic flame temperature ($T_b(H_2-air) \approx 2100K$) and larger density of the gas mixture the radiative preheating is only approximately 1.3 times greater than for the hydrogen-oxygen, however the flame velocity increased in this case considerably - 2.5 times.

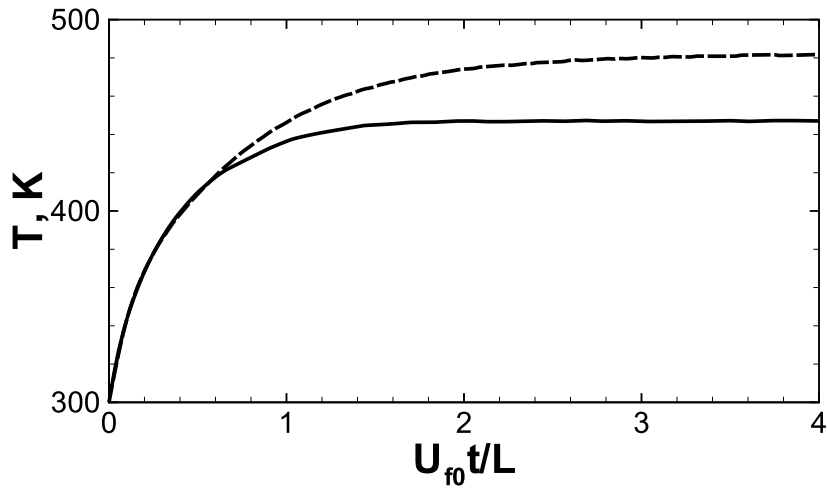

**Figure 4.** Temporal evolution of the gaseous temperature during radiative preheating at the distance 2mm ahead the flame front for $H_2/O_2$ (solid line) and $H_2$/air (dashed line) flames for radiation absorption lengths, L=1cm. On the x-axis time is in units $L/U_{f0}$.

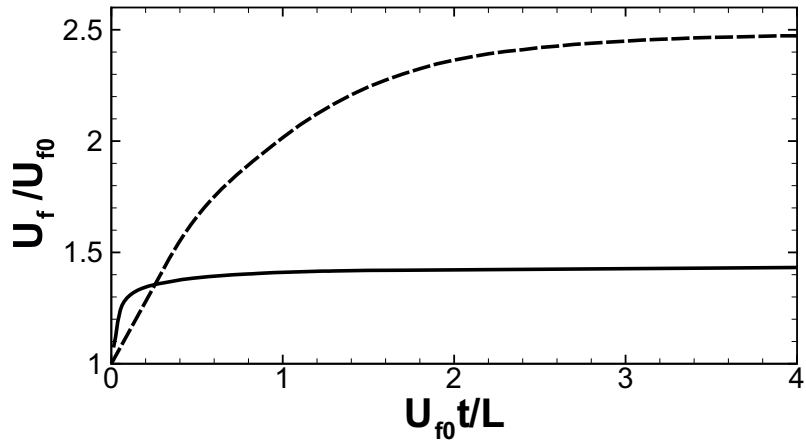

**Figure 5.** Temporal evolution of the $H_2/O_2$ (dashed line) and $H_2$/air (solid line) flame velocity for the conditions in Fig.4. The flame velocities are normalized on the corresponding normal laminar flame velocity $U_{f0}$ in pure gas mixture. Time is in units $L/U_{f0}$.

Fig. 4 shows the time evolution of the gaseous temperature ahead the flame front for $H_2$-$O_2$ and $H_2$-air flames for the radiation absorption length, L=1cm. The corresponding increase of the flame velocity relative to the unburned mixture due to the radiation preheating shown in Fig.4 for $H_2$-air flame and for $H_2$-$O_2$ flame is presented in Fig.5.

For the thermal radiation heat transfer to become a dominating process the necessary condition is that the characteristic time of ignition ahead of the flame to be small compared with the time the original flame travels through the radiation preheat zone, $\tau_{ind}(T_{cr}) \ll L/U_f$. It is unlikely that this condition can held for fast $H_2/O_2$ and $H_2$/air flames, but it is likely possible for a very slow flame, such as e.g. methane/air flame.

All the same, in the case of non-uniform distribution of particles (shown schematically in Fig. 6), the time of the radiative heating can be long enough to rise temperature in the mixture ahead of the flame above the crossover value. The result of the radiation preheating is an inhomogeneous in space temperature distribution formed in the unburned mixture ahead of the flame with the steepness of the temperature gradient determined by the thermal radiation absorption length. If the maximum temperature within the temperature gradient established due to the radiative preheating of the gas-particle mixture exceeded the crossover value, then depending on the steepness of the temperature gradient either deflagration or detonation can be ignited via the Zeldovich's gradient mechanism [1, 2].

## 4. The radiative preheating of non-uniformly dispersed particles: Ignition of deflagration and detonation ahead of $H_2/O_2$ flame

Consider a non-uniform distribution of the particles shown schematically in Fig. 6. Concentration of the particles immediately ahead of the flame front (gap 2 in Fig.6) is relatively low so that the radiation absorption length here is much larger then the gap width. Below we assume that the "gap" between the flame and the left boundary of the particles cloud is transparent for the thermal radiation so that the radiant heat flux is absorbed only in the layer 3. If time of the flame arrival to the boundary of the layer 3 is long enough, so that temperature of the particles and surrounding mixture can rise up to the value suitable for ignition before the flame arrival, which is about 1ms for $H_2/O_2$ flame (corresponding width of the gap is about 1cm), then the maximum temperature ($T_g$ in Fig.6) within the temperature gradient established due to the radiative preheating exceeds the crossover value. What kind of combustion regime is ignited via the Zel'dovich's gradient mechanism in the denser dust cloud depends on the radiation absorption length and, correspondingly, the steepness of the formed temperature gradient. The ignition starts when the temperature exceeds the crossover value, which is for hydrogen/oxygen at 1atm is 1050±50K.

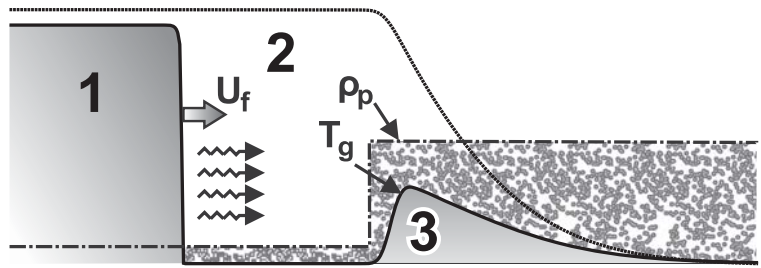

**Figure 6.** Scheme of the radiation preheating of the gaseous mixture inside the gas-particles cloud ahead the flame front. 1- high temperature combustion products; 2 - "gap" with lower concentration of particles; 3 –cloud of particles; $T_g$ -temperature of the radiative preheated gas.

The temperature gradient established due to the radiative preheating of the mixture in the gas-particle cloud depends mainly on the radiation absorption length and it is also influenced by the gas expansion during the heating. Since the characteristic acoustic time in the preheat zone is much smaller then time of the radiative heating up to crossover temperature, the pressure is equalized within the region heated by the radiation, and the temperature gradient is formed at the constant pressure $P \approx P_0 = 1atm$. Classification of the combustion regimes in hydrogen/oxygen and hydrogen/air mixtures initiated by initial temperature gradient via the Zel'dovich's gradient mechanism has been studied in [41, 42] using a detailed chemical kinetic models.

Figure 7 shows the calculated temporal evolution of the gaseous temperature and the particle mass density profiles during the preheating and establishing of the final temperature gradients at $t_0 \approx 900\mu s$, when the maximum temperature raised up to the crossover value. The calculations were performed for the initial stepwise density of particles, for particles of radius $r_p = 1\mu m$, with the maximum concentration of particles $N_p = 2.5 \cdot 10^7 cm^{-3}$ within the gas-particles layer. The scale $\Delta = (T^* - T_0)/|dT/dx| \approx 1 cm$ of the temperature gradients in Fig. 7, which are formed near the left boundary of the particle-gas cloud is close to the value of the radiation absorption length $L = 1/\pi r_p^2 N_p \approx 1.2 cm$.

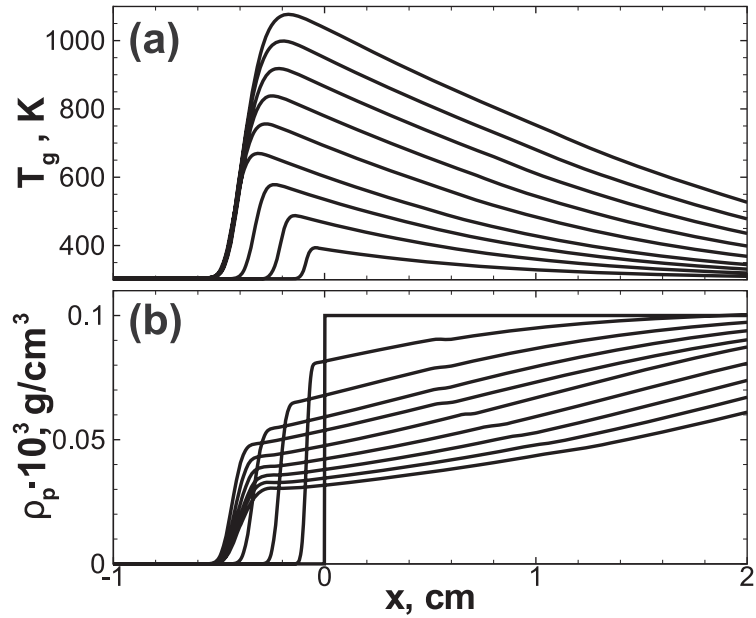

**Figure 7.** Time evolution of the gaseous temperature (a) and the mass density of the suspended solid particles (b) profiles during radiative preheating inside the gas-particles cloud ahead of the propagating flame. Profiles are shown with the time intervals of 50μs. For initial stepwise particles density profile, $N_p = 2.5 \cdot 10^7 cm^{-3}$, $r_p = 1\mu m$.

According to the classification of combustion regimes initiated by the initial temperature gradient in hydrogen oxygen at 1 atm [41, 42], such temperature gradient can ignite a deflagration. The time evolution of the gaseous temperature profiles presented in the middle frame of Fig. 8 shows development of the spontaneous reaction wave on the formed temperature gradient, which then transits into a deflagration wave. The dashed line in the upper frame of Fig. 8 shows the initial number density profile of the particles and the solid line shows the particles number density profile formed at the instant $t_0 = 900\mu s$ prior to the ignition, when the temperature gradient with maximum temperature $T^* = 1050\,K$ is formed (the first temperature profile in the middle frame). The calculated evolution of the pressure profiles in the bottom frame of Fig. 8 indicates a small variation of pressure during the formation of deflagration.

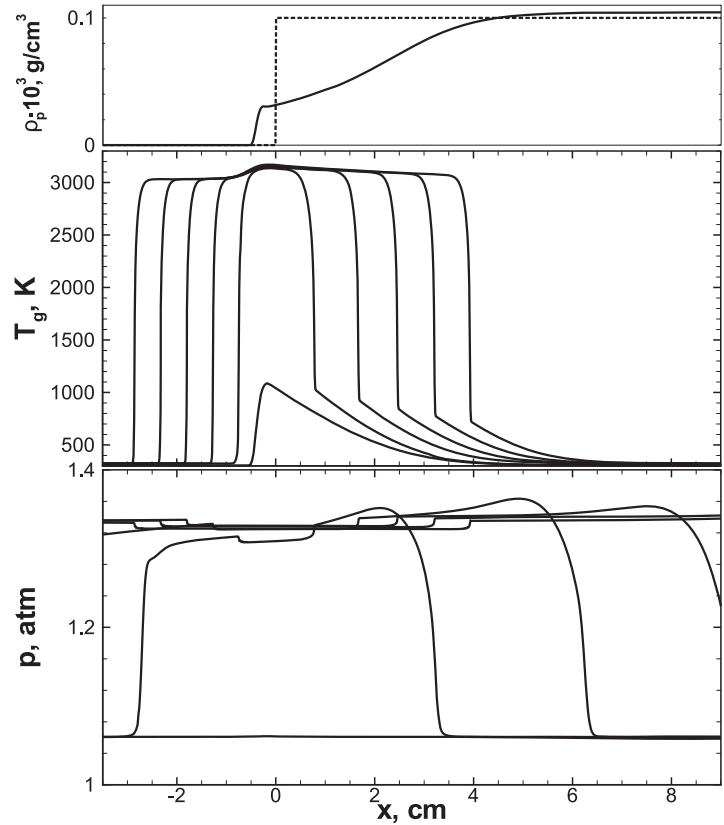

**Figure 8**. Temporal evolution of the gaseous temperature profiles (middle frame) and pressure profiles (bottom frame) during the slow combustion wave formation in the vicinity of the margin of the gas-particles cloud far ahead the propagating flame front, $t_0 = 900\mu s$, $\Delta\tau = 50\mu s$. The upper frame shows the distribution of particles mass density: the initial stepwise density profile (dashed line) and density profile at time instant $t_0$ prior to the ignition (solid line).

A more gentle temperature gradient can be formed either in the cloud with smaller concentration of the particles, or for the cloud with a properly diffuse interface instead of the stepwise particles number density distribution. In the latter case the radiation absorption length varies along the diffusive cloud interface resulting in the formation of a smooth temperature profile with a shallow temperature gradient capable to initiate either fast deflagration or detonation. Examples of the particle clouds with diffuse interface and the calculated temperature profiles caused by the radiative preheating are shown in Figs.9 and 10.

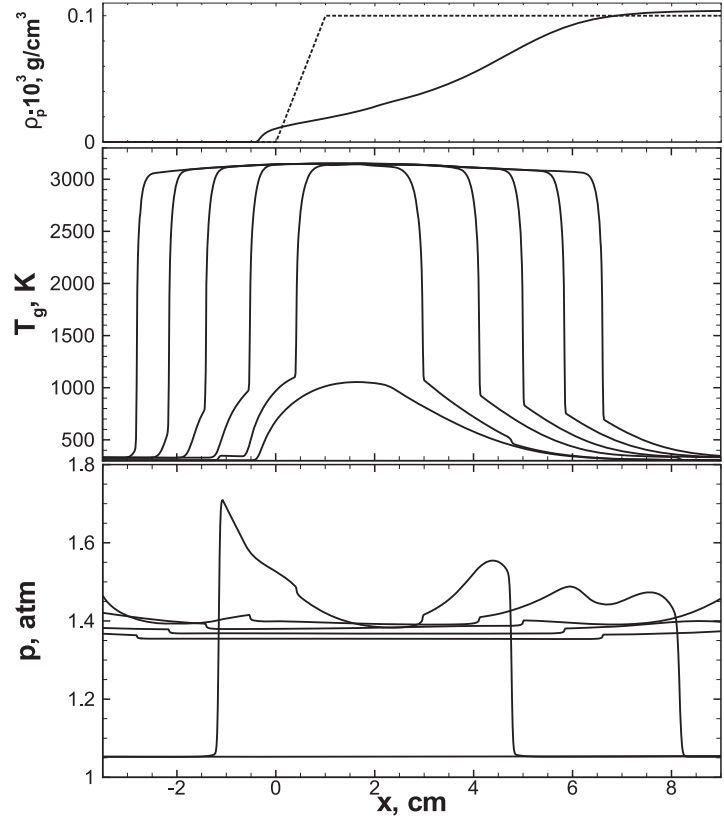

**Figure 9.** Temporal evolution of the gaseous temperature (middle frame), particles mass density (upper frame) and pressure (bottom frame) profiles during the fast combustion wave formation behind the outrunning shock in the vicinity of the margin of the gas-particles cloud ahead of the original propagating flame: $t_0 = 1650\mu s$, $\Delta\tau = 50\mu s$. The initial linear density profile of width 1.0cm (dashed line), density profile at $t_0$ prior to the ignition (solid line).

The upper frame in Fig. 9 shows the initial (dashed line) number density of the particles, which drops linearly on the scale 1cm from its maximum value $N_p = 2.5 \cdot 10^7 cm^{-3}$. The diffuse boundary of the particles cloud is smeared during the radiation preheating due to the expansion of the gas and at the instant $t_0 = 1600\mu s$ prior to the ignition, when the temperature gradient with maximum temperature $T^* = 1050 K$ is formed it is shown by solid line. The middle frame in Fig. 9 shows the calculated temporal evolution of the gaseous temperature profiles. It depicts the development of spontaneous reaction wave on the formed temperature gradient of the scale $\Delta = (T^* - T_0)/|dT/dx| \approx 8cm$, which then transits into the fast deflagration behind the outrunning shock in the vicinity of the gas-particles layer boundary. The weak shock waves outrunning the deflagration are seen in the bottom frame of Fig. 9.

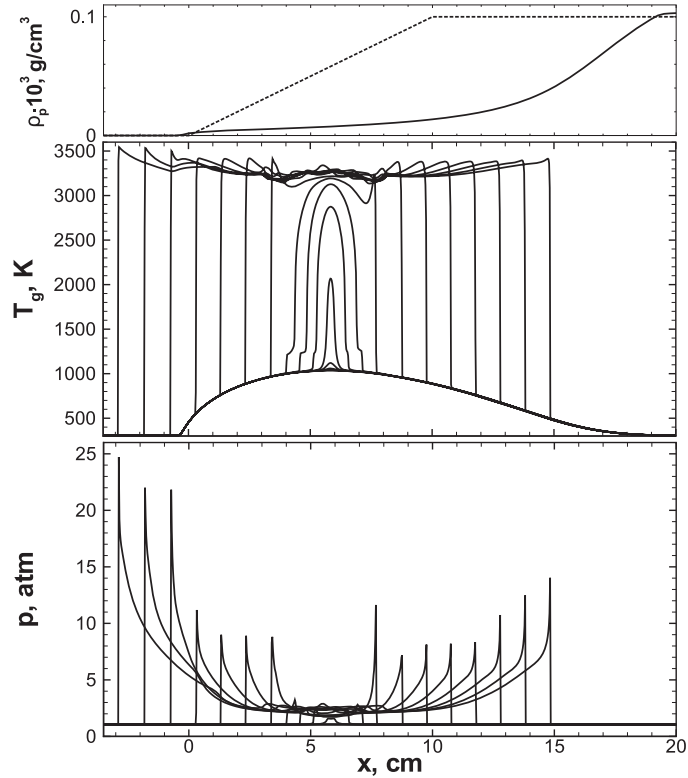

**Figure 10.** Temporal evolution of the gaseous temperature profiles (middle frame) and pressure profiles (bottom frame) during the detonation formation in the vicinity of the gas-particles cloud boundary ahead the propagating flame: $t_0 = 4980\mu s$, $\Delta \tau = 4\mu s$. The upper frame shows the distribution of particles mass density: the initial linear density profile of width 10.0cm (dashed line) and density profile at time instant $t_0 = 4980\mu s$ prior to the ignition (solid line).

The initiation of a detonation by the initial temperature gradient requires more shallow gradient. According to [41, 42] the minimum scale of the initial linear temperature gradient in hydrogen/oxygen mixture at normal conditions ($P_0 = 1atm$) and $T^* = 1050\,K$ at the top of the gradient needed for the detonation initiation is $(T^*-T_0)/\nabla T \approx 20cm$. Figure 10 presents results of the simulation for the layer with initial diffuse boundary with the particles number density, which drops linearly on the scale of 10cm from its maximum value $N_p = 2.5 \cdot 10^7 \text{cm}^{-3}$. The upper frame in Fig. 10 shows the initial particles number density (dashed line) and the formed diffuse boundary of the particles cloud smeared due to gas expansion during the thermal radiation at the instant $t_0 = 4980\mu s$ prior to the ignition (solid line). Temporal evolution of the gaseous temperature during the detonation formation in the vicinity of the diffusive layer boundary is depicted in the middle frame in Fig.10. The temporal evolution of the temperature profile corresponds to the development of spontaneous reaction wave, its coupling with the shock wave and the formation of the detonation wave. Temporal evolution of the pressure profiles, corresponding to the formation of shock wave, its coupling with the spontaneous reaction wave and formation of strong shock corresponding to a detonation wave is seen in the bottom frame of Fig. 10.

## 5. Discussion and conclusions

The present study demonstrates that the radiative heat transfer in a particle-laden gaseous flame can considerably influence the overall picture of the flame propagation. It is shown that depending on the spatial distribution of the suspended particles, the radiative preheating can considerably intensify the gaseous burning, leading to an increased flame velocity and can promote formation of the temperature gradients, which can trigger off new combustion regimes ahead of the primary flame via the Zel'dovich gradient mechanism with possible triggering of a detonation.

The performed numerical simulations demonstrate the plausibility of radiation preheating as the principal effect of the combustion intensification and in some cases initiation of detonation in the gaseous fuel, where relatively low concentration of suspended solid particles or any other substance can absorb the radiative heat flux and rise temperature of the fuel ahead of the flame. The presented results show that the thermal radiative preheating play a significant role in determining the regimes of combustion in two-phase reacting flows.

It should be emphasized that this study is a necessary prerequisite aiming to show principle physics and role of the radiative preheating, which can be important for understanding different combustion phenomena. The radiative preheating and the radiative heat transfer can be important for understanding different combustion phenomena at terrestrial conditions and in astrophysics. The conditions under which a reactive two-phase mixture can ignite and produce a heat release are important in different areas of fire safety. In particular, the preheating of the combustible mixture ahead of the flame, due to the absorption of thermal radiation emitted from the flame by the suspended particles, which results in flame acceleration and triggering of detonation, is a plausible rout in order to identify the nature of dust explosion. The danger of dust explosion exists in processes that are accompanied by the formation of clouds of fine dust particles. These events are common risks in the coal, metallurgy, chemicals, wood, hydrogen and hydrogen-based technologies and other industries. Recent experiments have shown that the dust cloud flame propagation is strongly influenced by the thermal radiation [24-27].

Triggering a detonation by the temperature gradient formed ahead of the flame due to the radiation preheating, considered in the present study, can be plausible scenario of the transition to detonation in the thermonuclear Type Ia supernovae (SN Ia) explosion. Type Ia supernovae have received increased interest because of their importance as 'standard candles' for cosmology. Observations using Type Ia supernovae as standard candles have revealed that the expansion rate of the universe is accelerating and have led to the discovery of dark energy [47-49]. Because of their extreme and predictable luminosity, SN Ia are extensively used as standard candles to measure distances and estimate cosmological parameters critical for our understanding of the global evolution of the Universe. To improve these measurements, we need comprehensive theoretical and numerical models of SN Ia that describe details of the explosion and connect them to observed characteristics of SN Ia, such as spectra and light curves. But the way in which Type Ia supernovae explode is not completely understood. The current leading paradigms for the explosion is scenario of the deflagration-to-detonation transition. There is increasing evidence that a detonation is needed to explain majority features of the SN Ia explosion. Nowadays the best modeling describing majority of the observed SN Ia events is provided by the so-called delayed detonation model [50-52], which imply a phase of subsonic thermonuclear burning (deflagration) during which the star expands and a phase of a detonation, which burns remaining fuel on timescales much shorter than the timescale of the explosion. The paradigm of the delayed detonation models is consistent with the theoretical models [53, 54] and with theoretical conclusion [55] that a detonation in a strongly Fermi-degenerated matter is unstable against 1D pulsations at densities higher than $2 \cdot 10^7 \, \text{g}/\text{cm}^3$ and it becomes stable at lower densities near the star surface. There has been numerous attempts both analytical and numerical [50-60] to explain the detonation formation in SN Ia explosion. However, after many years of studies a fundamental question what is the mechanism of deflagration-to-detonation transition (DDT) in the supernova Type Ia events still remains the least understood aspect of the SN Ia explosion phenomenon (see e.g. [59]).

Type Ia supernovae begins with a white dwarf (WD) near the Chandrasekhar mass that ignites a degenerate thermonuclear runaway close to its center and explodes having initial radius $R_{WD} \approx 10^8 \, \text{cm}$. Such a wide range of the length scales necessitates to use models of the infinitely thin flame (see [59] for a review of explosion scenarios) that limits sufficiently the understanding of the transient phenomena. In the context of thermonuclear burning of SN Ia, combustion initially proceeds in the deflagration mode from the center of SN Ia with the velocity and width of the laminar flame $U_f \approx 10^7 \, \text{cm}/\text{s}$ and $L_f \approx 10^{-2} \, \text{cm}$. Possible new mechanism of DDT in SN Ia explosion may be associated with detonation triggering due to radiation preheating similar to the scenario of the detonation ignition considered in Sec.4. During the late deflagration phase the radiant flux produced by the radioactive decays of $Ni^{56} \to Co^{56} \to Fe^{56}$ increases considerably in course of the star incineration by the expanding deflagration wave. Absorption of the radiant energy flux in the outer layers of the star may produce a preconditioned region with the shallow temperature gradient such that detonation can be ignited via the Zel'dovich gradient mechanism.

**Acknowledgements**

This work was supported in part by Ben-Gurion University Fellowship for senior visiting scientists (ML), by the Research Council of Norway under the FRINATEK, Grant 231444 (ML), and by the Program of Russian Academy of Sciences "Basic problems of mathematical modeling" (MI, AK). We acknowledge the allocation of computing resources provided by the Swedish National Allocations Committee at the Center for Parallel Computers at the Royal Institute of Technology in Stockholm, the National Supercomputer Centers in Linkoping and the Nordic Supercomputer Center in Reykjavik. One of the authors (M.L.) is grateful to Axel Brandenburg and Nils E. Haugen for useful discussions during writing this paper.

**Appendix A: Code validation, resolution and convergence tests**

The thorough convergence and resolution tests were carried out to verify that the observed phenomena of the radiation preheating and impact of the radiation heat transfer in particle-laden flame were correctly caught remaining unchanged with increasing resolution.

Figures A1 and A2 represent the results of the convergence test for a flame propagating through the pure gaseous stoichiometric hydrogen-oxygen mixture at normal conditions ($T_0 = 300K$, $P_0 = 1atm$) and for the mixture with suspended micro particles (for the uniform particles distribution, corresponding to Figs.1-3 ($L = 2cm$) in Section 3. In the latter case the combustion velocity and the flame structure is fully defined by the state of the gaseous mixture just ahead of the flame front. As the temperature in this region does not exceed 450÷500K and the pressure remains constant the convergence is for almost the same fine resolution as in the case for a pure gaseous mixture (see Fig. A1). As it was mentioned in Sec.3 in case of considerably smaller flame velocities the preheating is more efficient close to the flame front. In such conditions one should resolve flame structure at the higher temperatures and use finer meshes (see Fig. A2). The meshes were taken to resolve the structure of the flame front with 6, 8, 12, 24, and 48 computational cells, corresponding to the computational cell sizes: $\Delta$ = 0.1, 0.05, 0.025, 0.01 and 0.005mm, respectively. The acceptable quantitative convergence was found for resolution of 24 computational cells per flame front (see Figs. A1, A2). Therefore the resolution 24 and 48 computational cells was typically used for solving the problem in Section 3. As it is seen from Fig. A2, which shows the convergence tests for three different initial temperatures ($T_0 = 293$, 600, and 1000K, corresponding to curves 1, 2, 3 in Fig.3) the flame dynamics at the elevated temperatures can be resolved only with a finer resolution than at the lower temperatures. Therefore to obtain the converged solution for the detonation initiation problem one should use a finer resolution from the very beginning.

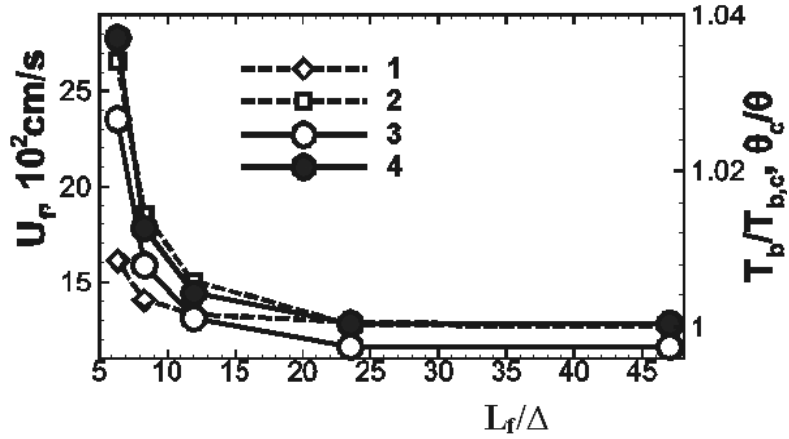

**Figure A1.** Resolution test for normal velocity of stoichiometric hydrogen-oxygen flame at normal ambient conditions ($T_0 = 300K$, $p_0 = 1atm$). $U_f$ – flame velocity reproduced with the computational cell of size $\Delta$; $T_b$ – adiabatic temperature of the combustion products; $\theta = \rho_u / \rho_b$ – expansion ratio, index 'c' corresponds to the converged values; Burning velocity is presented for two cases: without particles (empty signs) and for gaseous mixture with suspended particles (filled signs).

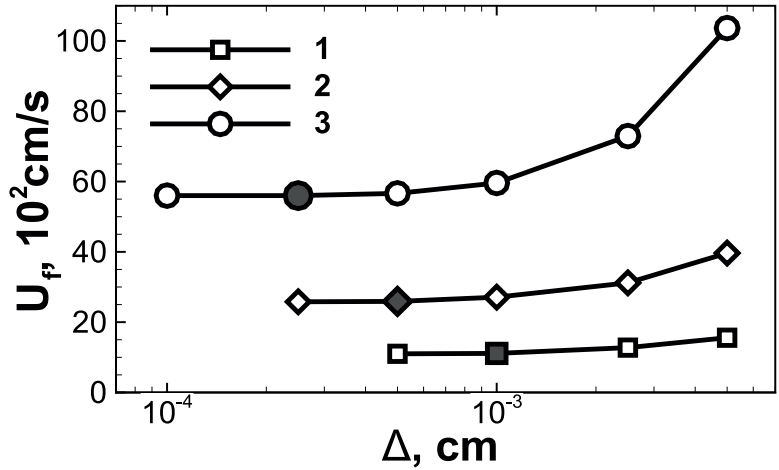

**Figure A2.** Resolution tests for different ambient temperatures (1- 293K; 2 - 600K; 3 - 1000K). Filled signs show acceptable range of convergence. $U_f$ – is a normal flame velocity reproduced with cell size $\Delta$.

The problem considered in Section 4 demands more robust parameters of the computational setup. The most demanding is the case where the detonation arises as a result of auto-ignition inside the hot-spot. The consequence of the processes taking place in such a case should be appropriately resolved. As it was shown in [41, 42] the detonation caused by the Zel'dovich gradient mechanism arose via the following scenario: 1) the spontaneous combustion wave were formed, 2) as the spontaneous wave decelerated the pressure wave formed behind its front overran it forming the shock wave, 3) detonation established after the transient process involving flame acceleration in the flow behind the outgoing shock wave.

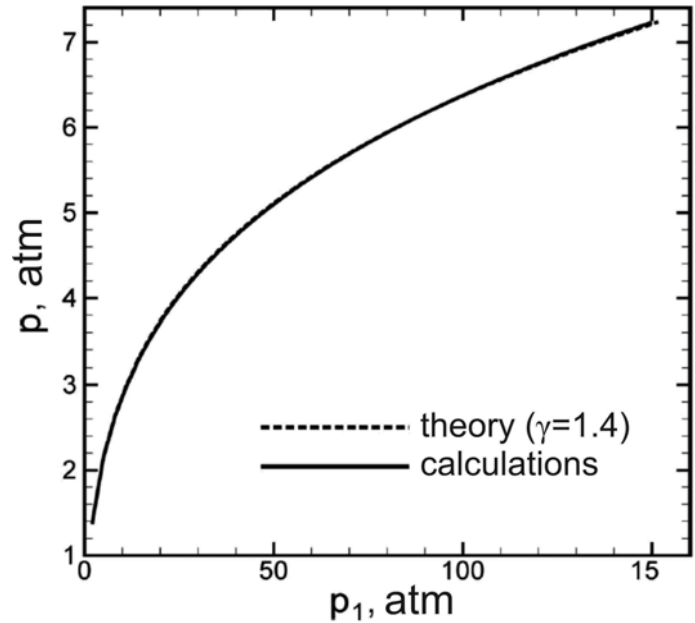

**Figure A3.** Comparison of calculated data (solid line) with analytical solution (dashed line) for discontinuity decay in air for different driver pressures ($p_1$) at normal ambient conditions ($p_0$=1atm, $T_0$=300K).

Due to this sequence of the events one should appropriately resolve the combustion waves propagating through reacting medium at initially elevated temperature (~1000K, see figures in Section 4) and on the background of elevated temperature and pressure behind the shock front. Besides, the coupling of the reaction wave and shock should be resolved taking into account that the flame thickness is much larger than the width of the shock front. According to this we performed a number of resolution tests for the flames propagating at elevated temperatures and pressures and interacting with the compression and shock waves that can arise in the compressible flow. Therefore to obtain the converged solution for the detonation initiation problem one should use a finer resolution from the very beginning. The hydrogen-oxygen flame thickness at normal conditions is about 0.24mm and one should resolve it with not less than 20 cells to avoid unphysical couplings with the shocks that are usually are smoothed over 5-6 cells even by the high-order schemes with limiters. This means that one should resolve flame structure at the higher temperatures and use finer meshes, as it is seen in Fig. A2, which shows resolution tests for different ambient temperatures with grey signs for acceptable range of convergence. We also additionally performed the tests of how shocks are reproduced with chosen numerical method. For example, Fig. A3 shows quantitative reliability of the computational method, demonstrating the comparison of calculated data (solid line) with analytical solution (dashed line) for discontinuity decay in air for different driver pressures ($p_1$) at normal ambient conditions ($p_0$=1atm, $T_0$=300K).

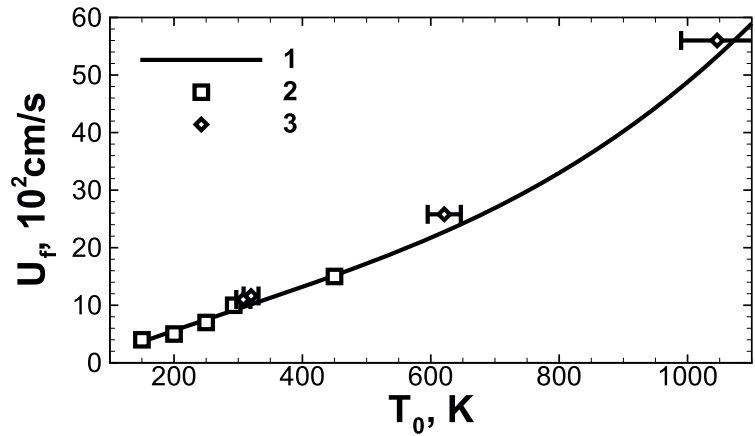

**Figure A4.** Stoichiometric hydrogen-oxygen burning velocities versus the ambient temperatures. Solid line shows the extrapolation of the experimental data obtained in [57] and signed with squares. Diamonds represent the calculated data with error bars determined by the temperature rise behind the compression wave emerged from the ignition kernel.

Figure A4 shows the burning velocity-temperature dependence. One can observe that agreement between calculated data and the extrapolation of the experimental data using the global reaction order $n = 2.74$ is satisfactory good. The error bars for calculations are due to the lack of information about the ignition conditions. The ignition conditions in the calculations were such that the compression wave arose and outran from the ignition zone preheating the mixture ahead the forming flame front that could cause the discrepancy with the experimental data [61]. For the model considered in the present paper the resolution was taken with at least 48 computational cells per flame front at normal conditions. It agrees well with the results obtained previously in [41, 42]. The performed resolution and convergence tests have shown that the processes in the considered problem setup are well resolved with the chosen method and chosen grids. There is almost no influence of scheme diffusion and all the scales and values are reproduced with high enough accuracy.